\documentclass[12pt,letterpaper]{article}
\pdfoutput=1

\usepackage{verbatim}

\usepackage{amsmath,amsfonts,mathrsfs,mathtools}
\usepackage{times}

\usepackage{graphics,xcolor}

\usepackage[authoryear]{natbib}

\usepackage{enumerate}

\usepackage{psfrag,pstricks,pst-all}
\usepackage[subrefformat=parens,labelformat=parens]{subfig}

\usepackage{url}

\newcommand{\gam}{\mathrm{Gam}}
\newcommand{\bet}{\mathrm{Beta}}

\newcommand{\expo}{\mathrm{Exp}}
\newcommand{\E}{\mathrm{E}}

\newcommand{\transpose}{^\mathrm{T}}
\renewcommand{\mid}{|}

\newcommand{\vect}[1]{\boldsymbol{#1}}
\newcommand{\matr}[1]{\boldsymbol{#1}}

\bibpunct{(}{)}{;}{a}{}{,}  

\begin{document}
\begin{flushright}
Version dated: \today
\end{flushright}
\bigskip

\bigskip
\medskip
\begin{center}

\noindent{\Large \bf Generalising rate heterogeneity across sites in statistical phylogenetics} 
\bigskip

\noindent {\normalsize \sc Sarah E.~Heaps$^1$, Tom M.~W.~Nye$^1$, Richard J.~Boys$^1$, \\
Tom A.~Williams$^2$, Svetlana Cherlin$^3$ and T.~Martin Embley$^4$}
\bigskip

\noindent {\small \it 
$^1$School of Mathematics, Statistics and Physics, Newcastle University, Newcastle upon Tyne, NE1 7RU, U.K.\\ 
$^2$School of Biological Sciences, University of Bristol, Bristol, BS8 1RJ, U.K.\\
$^3$Institute of Genetic Medicine, Newcastle University, Newcastle upon Tyne, NE1 3BZ, U.K.\\
$^4$Institute for Cell and Molecular Biosciences, Newcastle University, Newcastle upon Tyne, NE2 4HH, U.K}.\\
\end{center}
\bigskip


\subsubsection*{Abstract}
Phylogenetics uses alignments of molecular sequence data to learn about evolutionary trees relating species. Along branches, sequence evolution is modelled using a continuous-time Markov process characterised by an instantaneous rate matrix. Early models assumed the same rate matrix governed substitutions at all sites of the alignment, ignoring variation in evolutionary pressures. Substantial improvements in phylogenetic inference and model fit were achieved by augmenting these models with multiplicative random effects that describe the result of variation in selective constraints and allow sites to evolve at different rates which linearly scale a baseline rate matrix. Motivated by this pioneering work, we consider an extension using a quadratic, rather than linear, transformation. The resulting models allow for variation in the selective coefficients of different types of point mutation at a site in addition to variation in selective constraints.

We derive properties of the extended models. For certain non-stationary processes, the extension gives a model that allows variation in sequence composition both across sites and taxa. We adopt a Bayesian approach, describe an MCMC algorithm for posterior inference and provide software. Our quadratic models are applied to alignments spanning the tree of life and compared with site-homogeneous and linear models.

\bigskip

\noindent (\textbf{Keywords:} across-site rate heterogeneity; compositional heterogeneity; multiplicative random effects; phylogenetics; selective coefficients; tree of life.)


\section{\label{sec:intro}Introduction}
In statistical phylogenetics, the goal is to learn about the evolutionary relationships amongst a collection of species, generally using DNA or protein sequence data. These relationships are represented through a rooted, bifurcating tree called a phylogeny. Substitutions in the molecular sequence alignment are typically modelled using continuous time Markov processes, parameterised through an instantaneous rate matrix. Early phylogenetic models were simplistic, generally assuming that the evolutionary process was in its stationary distribution and that substitutions at each site of the alignment could be described by the same underlying rate matrix. Under these models, the probability of change from one character state to another was therefore independent of both organismal lineage and the biochemical function of the nucleotide or amino acid in question. These simplifying assumptions were known to be false, but were made for the sake of mathematical convenience and computational tractability, given the limited computing power for model fitting available at the time. In particular, it was already clear to early molecular evolutionists that rates of evolution vary according to functional or structural pressures acting at a site: important sites are subject to high selective constraints and evolve slowly because most mutations that arise at those sites are eliminated from the population by negative selection \citep{FM70}. \citet{UZ71} showed that the numbers of substitutions occurring at different sites could be modelled using a negative binomial distribution. Later, \citet{Yan93} incorporated the idea into statistical phylogenetics by allowing different sites to evolve at different rates. These rate parameters scaled the underlying Markov process rate matrix and were modelled as multiplicative random effects, with unit mean gamma distribution.

Incorporation of across-site rate variation into standard, stationary substitution models has led to major improvements in model fit and to the accuracy of phylogenetic inference \citep[][]{Yan96}. But there are other, pervasive features of molecular sequence data that these models do not accommodate. In particular, nucleotide composition is believed to vary across \emph{both} sites of the alignment \emph{and} branches of the phylogenetic tree. For example, the GC-content of ribosomal DNA genes varies from 45-74\% across the known diversity of cellular life \citep[][]{CFHHE08}, implying that the probabilities of each of the four nucleotides can change over time. These compositional shifts might reflect changing biases in DNA repair enzymes \citep[][]{Sue88} or, at least for genes encoding structural RNAs, adaptation to different growth temperatures \citep[][]{GL97}. As well as variation in sequence composition across taxa, there is also compositional variation observed among the different sites within an individual protein-coding sequence: due to functional constraints, most sites can tolerate only a limited, and typically biochemically homogeneous, subset of the twenty amino acids \citep[][]{FM70}. The result is that, in addition to varying in evolutionary rate, sites can also differ in sequence composition. As with heterogeneity in evolutionary rates, failure to account for variation in composition can lead to model misspecification and, therefore, serious phylogenetic error, as demonstrated by a number of empirical studies \citep[][]{ETW93,Fos04,LBP07,Phi11}. The phylogenetic literature includes a number of models designed to capture one type of compositional heterogeneity, or the other, that is, \emph{either} heterogeneity across sites \emph{or} heterogeneity across branches. In the former case, this is often achieved using mixture models which classify sites into groups, each of which has a different stationary distribution; see, for example, \citet{PMC04} or \citet{LP04}. To allow heterogeneity across branches, a number of models have been developed which drive the Markov process towards a different stationary distribution at different points on the tree, typically by allowing evolution on different branches to be governed by different instantaneous rate matrices; see, for example, \citet{YR95}, \label{pg:rev3_4a}\citet{BL06}, \citet{DB08} or \citet{HNBWE14}. Despite the large body of literature focused on modelling compositional heterogeneity of one type or the other, there have been very few attempts to model both jointly. Such efforts are typically based on mechanistic models which allow different rate matrices to govern the evolutionary process on different (site, branch) pairs; see, for example \label{pg:rev3_4b}\citet{BL08} or \citet{JWRPJ14}. Unfortunately, use of these models has been limited due to computational difficulties with model-fitting.

In a simple phylogenetic model, evolution at all sites is controlled by a single instantaneous rate matrix. The across-site rate variation model offers greater flexibility by allowing site-specific \emph{linear} transformations of the baseline rate matrix, with variation amongst scaling factors dependent on a single-parameter gamma distribution. Owing to the success of this simple modification, the across-site rate variation model has been extended in a number of ways. \label{pg:rev1_1b}For example, covarion models \citep[][]{TS98,Hue02,Gal01} allow the site-specific (linear) scaling factors to vary from branch to branch. This is intended to capture the variation over time in selective constraints that arise as a consequence of earlier substitutions at other sites. In this paper we consider a different generalisation of the across-site rate variation model, applying site-specific \emph{quadratic}, rather than linear, transformations of a baseline matrix. This gives a more flexible model which is dependent on an additional unknown parameter. \label{pg:rev3_1i}It has the effect of allowing variation in the selective coefficients -- that is, the strength of selection -- for different types of point mutation at a site, in addition to heterogeneity in the overall selective constraints across sites. We thereby obtain a more biologically plausible model. \label{pg:rev2_1a}Further, we demonstrate that when linear or quadratic across-site transformations are combined with a class of non-stationary Markov processes, we obtain computationally tractable models that allow sequence composition to vary across \emph{both} branches of the tree \emph{and} sites of the alignment, addressing the clear need in the literature for models of this type. 

The remainder of this paper is organised as follows. Section~\ref{sec:models} introduces phylogenetic models of sequence evolution and the incorporation of multiplicative random effects to allow rate variation across sites. Section~\ref{sec:quash} describes our quadratic generalisation and its properties. In Section~\ref{sec:ns-quash} we combine across-site linear and quadratic transformations with a general class of non-stationary substitution models and describe the properties of the resulting Markov processes. Section~\ref{sec:inference} addresses the issue of inference for models incorporating our quadratic transformation. Specifically, working in a Bayesian framework, we describe the posterior distribution of interest and details of our numerical approach to model-fitting via Markov chain Monte Carlo sampling. In Section~\ref{sec:apps} we consider analyses of two biological data sets; the first involving a stationary model and the second, a non-stationary model. In each case we compare the performance of a site-homogeneous model with analogous models incorporating linear and quadratic across-site transformations of the baseline rate matrix. Finally, we summarise our conclusions in Section~\ref{sec:discussion}.

\section{\label{sec:models}Phylogenetic models of sequence evolution}
Denote by $\matr{y} = (y_{i,j})$ an alignment of molecular sequence data where $y_{i,j} \in \Omega_K$ is the character at the $j$-th site for taxon $i$ and $\Omega_K$ is an alphabet with $K$ characters, for example, the DNA alphabet with $\Omega_4 = \{ {\tt A}, {\tt G}, {\tt C}, {\tt T} \}$. Denote the number of sites (columns) by $M$ and the number of taxa (rows) by $N$ and let $\vect{y}_j = (y_{1,j},\ldots,y_{N,j})\transpose$ be the $j$-th column in the alignment. Consider a rooted, bifurcating tree $\tau$, with branch lengths $\vect{\ell}$, representing the evolutionary relationships amongst this collection of $N$ taxa. For every site, phylogenetic models typically assume that evolution along each branch of the tree can be modelled using a continuous time Markov process $Y(t)$, characterised by an instantaneous rate matrix $\matr{Q}=(q_{u,v})$ which has positive off-diagonal elements and rows that sum to zero. \label{pg:rev3_5a}This matrix controls the dynamics of the substitution process through the matrix equation $\matr{P}(\ell) = \{ p_{u,v}(\ell) \} = \exp (\ell \matr{Q}')$, where $\matr{Q}' = \matr{Q} / (-\sum_u q_{uu} \pi_u)$ and $\vect{\pi} = (\pi_1,\ldots,\pi_K) \in \mathscr{S}_K$ is the stationary distribution of the process. The notation $\mathscr{S}_K = \{(x_{1}, \ldots, x_{K}): x_i \geq 0 \; \forall \, i, \sum x_i = 1\}$ denotes the $K$-dimensional simplex. This rescaling of the rate matrix $\matr{Q}$ allows the branch lengths $\ell$ to be in interpreted as the expected number of substitutions per site. The $(u,v)$-th element in the transition matrix $p_{u,v}(\ell) = \Pr(Y(\ell) = v \mid Y(0) = u)$ for $u,v=1,\ldots,K$ is the probability of transitioning from character $u$ to character $v$ along a branch of length $\ell$.

Standard phylogenetic models assume that the underlying continuous time Markov process is time reversible and in its stationary distribution $\vect{\pi}$. Reversibility implies that $\pi_{u} p_{u,v}(\ell) = \pi_{v} p_{v,u}(\ell)$ for all $u,v$ and allows the rate matrix to be represented in the form $\matr{Q} = \matr{S} \matr{\Pi}$, where $\matr{\Pi} = \text{diag}(\vect{\pi})$, and $\matr{S}$ is a symmetric matrix whose off-diagonal elements, $\rho_{u,v}$ with $\rho_{u,v}=\rho_{v,u}$, are termed \emph{exchangeability} parameters. The latter determine the general propensity for change
between the different pairs of characters. We define a rate matrix as reversible if it permits a parameterisation of this form. The most general reversible rate matrix, with $K(K-1)/2$ distinct exchangeabilities, characterises the general time-reversible (GTR) model. Other commonly used substitution models are special cases. For example, the TN93 model is a special case for nucleotide data where $\rho_{C,T} = \rho_{T,C} = \rho_1$, $\rho_{G,A} = \rho_{A,G} = \rho_2$ and all other $\rho_{u,v}$ are equal to $\beta$. This simplification reduces the number of exchangeabilities from six to three but retains biological realism by allowing transversions (substitutions between a pyrimidine and a purine) and the two types of transitions (substitutions between pyrimidines and between purines) to occur at different rates, here $\beta$, $\rho_1$ and $\rho_2$ respectively.

Classically, the sites of the alignment $\matr{y}$ are assumed to evolve independently of each other and so the likelihood is given by
\begin{equation*}
p(\matr{y} | \matr{Q}, \tau, \vect{\ell}) = \prod_{j=1}^M \Pr(\vect{Y}_j = \vect{y}_j | \matr{Q}, \tau, \vect{\ell}).
\end{equation*}
\label{pg:rev3_5b}In order to prevent arbitrary rescaling of the rate matrix $\matr{Q}$ in the transition matrix $\matr{P}(\ell) = \exp (\ell \matr{Q}')$, where $\matr{Q}' = \matr{Q} / (-\sum_u q_{uu} \pi_u)$, it is common to impose an identifiability constraint, \label{pg:rev3_17a}for example by assuming that the exchangeability parameters sum to one or by fixing one of the exchangeability parameters $\rho_{u,v}$, $u \ne v$, to be equal to one \citep[][]{ZH04}. For instance, one can fix $\beta=1$ in the TN93 model. This allows the remaining exchangeability parameters to be interpreted as relative rates of change. We take the latter approach in this paper. Henceforth, we drop the prime on the normalised rate matrix $\matr{Q}'$ for notational brevity.

\subsection{\label{subsec:asrh}Modelling rate heterogeneity across sites}
It has long been recognised that selective pressures vary across sites
due to their differing roles in the structure and function of the
molecular sequence \citep[][]{Yan96,SNS96}. This feature is typically captured by a simple modelling device that allows
each site $j$ to evolve at its own rate $c_j > 0$ which scales the normalised rate matrix $\matr{Q}$ linearly. To enable information to be shared between
sites, the rates $\vect{c}=(c_1,\ldots,c_M)\transpose$ are generally
assumed to follow a gamma distribution with unit mean. Defining
\begin{equation}\label{eq:Q_j_lin}
\matr{Q}_j = c_j \matr{Q},
\end{equation}
the likelihood can then be represented as
\begin{equation*}
p(\matr{y} | \matr{Q}, \tau, \vect{\ell}, \alpha) = \prod_{j=1}^M \int_0^{\infty} p(c_j \mid \alpha) \Pr(\vect{Y}_j = \vect{y}_j | \matr{Q}_j, \tau, \vect{\ell}) \, d c_j,
\end{equation*}
where $p(c_j \mid \alpha)$ is the $\mathrm{Gam}(\alpha, \alpha)$ density function evaluated at $c_j$. The single parameter $\alpha$ determines the manner and extent to which the scaling factors differ across sites. We refer to models in which a baseline rate matrix is transformed according to~\eqref{eq:Q_j_lin} as \textit{linear across-site heterogeneity} (LASH) models. In order to simplify computation, the (continuous) gamma density $p(c_j \mid \alpha)$ is typically replaced by a discrete approximation with $K_c$ categories, most often $K_c=4$ \citep[][]{Yan94}. In a Bayesian setting, this numerical integration strategy may seem less natural than using data augmentation during MCMC and sampling the $c_j$. However, the discretisation allows much more caching of intermediate likelihood calculations which can substantially speed up computational inference.

In this model, the rate matrix at each site is simply a linearly scaled version of some underlying normalised baseline $\matr{Q}$. The transformation does not affect the theoretical stationary distribution, defined as the solution of $\vect{\pi} \matr{Q} = \vect{0}\transpose$, or, in the class of reversible models, the ratios of the exchangeability parameters. In the following section we generalise this model to allow the rate matrix at each site to be a more flexible \emph{quadratic} function of the base matrix, which depends on the values of \emph{two} parameters. This transformation preserves the stationary distribution but allows the rankings of the instantaneous rates of change to vary between sites. The resulting model can be interpreted biologically as one which allows variation in the selective coefficients of different types of point mutation at a site, in addition to variation in the overall selective constraints across sites.

\section{\label{sec:quash}Quadratic across-site heterogeneity models}
Consider a baseline normalised rate matrix $\matr{Q}$. At site $j$, the instantaneous rate matrix $\matr{Q}_j = (q_{j,u,v})$ is given by
\begin{equation}\label{eq:Q_j}
\matr{Q}_j = c_j \matr{Q} - c_j d_j \matr{Q}^2
\end{equation}
where $c_j \in (0, \infty)$ and $d_j \in (l(\matr{Q}), u(\matr{Q}))$, which reduces to the simple LASH model when $d_j=0$. We call any model in which a baseline rate matrix is transformed in this way a \textit{quadratic across-site heterogeneity} (QuASH) model. The limits $l(\matr{Q})$ and $u(\matr{Q})$ depend on $\matr{Q}$ and ensure that $\matr{Q}_j$ is a valid rate matrix, that is
\emph{(i)} all off-diagonal elements are positive: $q_{j,u,v} > 0$, $\forall u \ne v$;
\emph{(ii)} all row sums are zero: $\sum_{v} q_{j,u,v} = 0$ $\forall u$.

Property \emph{(ii)} is automatically satisfied for any $d_j \in \mathbb{R}$. The proof is as follows. The $(u,v)$-th element of $\matr{Q}_j$ is given by
\begin{equation*}
q_{j,u,v} = c_j \left(q_{u,v} - d_j \sum_{w} q_{u,w} q_{w,v} \right).
\end{equation*}
Therefore the sum of the elements on row $u$ of $\matr{Q}_j$, $\sum_{v} q_{j,u,v}$, is given by
\begin{equation*}
c_j \left( \sum_{v} q_{u,v} - d_j \sum_{v} \sum_{w} q_{u,w} q_{w,v} \right) = 
c_j \left( 0 - d_j \sum_{w} q_{u,w} \sum_{v} q_{w,v} \right) = 
c_j (0 - d_j \times 0) = 0
\end{equation*}
for any $d_j \in \mathbb{R}$.

For property \emph{(i)} to be satisfied we need
\begin{equation}\label{eq:lower}
l(\matr{Q}) = \max\{ \mathcal{L}(\matr{Q}) \},  \quad \mathcal{L}(\matr{Q}) = \left\{ \frac{q_{u,v}}{\sum_{w} q_{u,w} q_{w,v}} \, : u \ne v \text{ and } \sum_{w} q_{u,w} q_{w,v} < 0 \right\}
\end{equation}
and
\begin{equation}\label{eq:upper}
u(\matr{Q}) = \min\{ \{ \infty \} \cap \mathcal{U}(\matr{Q}) \}, \quad \mathcal{U}(\matr{Q}) = \left\{ \frac{q_{u,v}}{\sum_{w} q_{u,w} q_{w,v}} \, : u \ne v \text{ and } \sum_{w} q_{u,w} q_{w,v} > 0\right\}.
\end{equation}
By definition, $l(\matr{Q}) \le 0$ and $u(\matr{Q}) \ge 0$. Note that the set $\mathcal{L}(\matr{Q})$ cannot be empty, $\mathcal{L}(\matr{Q}) \ne \emptyset$. To prove this, suppose $q_{a,b}$ is the largest off-diagonal element in $\matr{Q}$. Now
\begin{align*}
\sum_{w} q_{a,w} q_{w,b} &= q_{a,a} q_{a,b} + q_{a,b} q_{b,b} + \sum_{w \ne a,b} q_{a,w} q_{w,b}\\
&= - q_{a,b} \sum_{w \ne a} q_{a,w} +  q_{a,b} q_{b,b} + \sum_{w \ne a,b} q_{a,w} q_{w,b}\\
&= - q_{a,b} \sum_{w \ne a,b} q_{a,w} - q_{a,b}^2 + q_{a,b} q_{b,b} + \sum_{w \ne a,b} q_{a,w} q_{w,b}.
\end{align*}
However, $q_{w,b} < q_{a,b}$ for all $w \ne a$ and so
\begin{equation*}
\sum_{w \ne a,b} q_{a,w} q_{w,b} < q_{a,b} \sum_{w \ne a,b} q_{a,w}.
\end{equation*}
Because $- q_{a,b}^2$ and $q_{a,b} q_{b,b}$ are strictly negative it follows that
\begin{equation*}
\sum_{w} q_{a,w} q_{w,b} = - q_{a,b} \sum_{w \ne a,b} q_{a,w} + \sum_{w \ne a,b} q_{a,w} q_{w,b} - q_{a,b}^2 + q_{a,b} q_{b,b} < 0.
\end{equation*}

In contrast, the set $\mathcal{U}(\matr{Q})$ can be empty. Consider, for example, the normalised rate matrix of the Jukes Cantor model, all of whose off-diagonal elements are equal to $1/3$. In this case, $\sum_{w} q_{u,w} q_{w,v} = -4 / 9 < 0$ for all pairs $(u,v)$ with $u \ne v$. Therefore $l(\matr{Q}) = - 3 / 4$ whilst the upper limit $u(\matr{Q})$ is infinite.

To allow information to be shared between sites, we continue to assume that the coefficients $\vect{c}=(c_1,\ldots,c_M)\transpose$ of the linear term are conditionally independent and identically distributed (i.i.d.) with $c_j | \alpha \sim \gam(\alpha, \alpha)$ for some unknown hyperparameter $\alpha$. In an analogous fashion, we assume that the coefficients $\vect{d}=(d_1,\ldots,d_M)\transpose$ of the second order term are independent of $\vect{c}$ and conditionally i.i.d. with $d_j | \matr{Q}, \beta \sim \mathcal{F}(\beta)$ for some unknown $\beta$, where the form of the distribution $\mathcal{F}$ will be discussed in Section~\ref{subsec:random_effects}. The likelihood can then be represented as
\begin{equation*}
p(\matr{y} | \matr{Q}, \tau, \vect{\ell}, \alpha, \beta) = \prod_{j=1}^M \int_0^{\infty} \int_{l(\matr{Q})}^{u(\matr{Q})} p(c_j \mid \alpha) p(d_j \mid \matr{Q}, \beta) \Pr(\vect{Y}_j = \vect{y}_j | \matr{Q}_j, \tau, \vect{\ell}) \, d c_j \, d d_j
\end{equation*}
where $\matr{Q}_j$ was defined in~\eqref{eq:Q_j}. As with the simpler LASH model, substantial gains in computational efficiency can be achieved by replacing the continuous densities $p(c_j \mid \alpha)$ and $p(d_j \mid \matr{Q}, \beta)$ by discrete approximations with $K_c$ and $K_d$ categories, respectively. We choose to place point masses of probability $1/(K_c K_d)$ at locations $\{ z_{c,a}(\alpha), z_{d,a'}(\matr{Q},\beta) \}$ for $a = 1, \ldots, K_c$, $a' = 1, \ldots, K_d$ where $z_{c,a}(\alpha)$ is the $(a - 0.5)/K_c$ quantile in the distribution of $c_j | \alpha$ and $z_{d,a'}(\matr{Q},\beta)$ is the $(a' - 0.5)/K_d$ quantile in the distribution of $d_j \mid \matr{Q}, \beta$. The likelihood then simplifies to
\begin{equation}\label{eq:likelihood_quash1}
p(\matr{y} | \matr{Q}, \tau, \vect{\ell}, \alpha, \beta) \simeq \prod_{j=1}^M \frac{1}{K_c K_d} \sum_{a=1}^{K_c} \sum_{a'=1}^{K_d} \Pr\left[ \vect{Y}_j = \vect{y}_j | \matr{Q}_j\left\{ z_{c,a}(\alpha), z_{d,a'}(\matr{Q},\beta), \matr{Q} \right\}, \tau, \vect{\ell} \right].
\end{equation}

\subsection{\label{subsec:quash_props}Properties of QuASH Models}
It can easily be shown that the stationary distribution of $\matr{Q}_j = c_j \matr{Q} - c_j d_j \matr{Q}^2$ is the same as that of $\matr{Q}$; see 
Section~A.1 of our Online Supplementary Materials 
for a proof. Of course the same is also true under the simple linear scaling, $\matr{Q}_j = c_j \matr{Q}$, which we recover when $d_j = 0$.  In the latter case, the linear mapping can simply be regarded as a site-specific scaling of the branch lengths. In contrast, our quadratic transformation does not preserve the ratios of the instantaneous rates of change in the baseline rate matrix, allowing different patterns of substitution at different sites. This idea is most readily exemplified in the context of reversible models where the transformation results in a site-heterogeneous model in which the exchangeability parameters vary across sites. Elucidating further,
it is straightforward to show that if $\matr{Q}$ is reversible, then so is $\matr{Q}_j$; see 
Section~A.2 of our Online Supplementary Materials 
for a proof. It follows that the set of GTR rate matrices is closed under our quadratic transformation. This is also true for some special cases of the GTR rate matrix including the TN93 rate matrix which was introduced in Section~\ref{sec:models}. In this case, suppose that $\beta$, $\rho_1$ and $\rho_2$ are the transversion and transition rates in the baseline rate matrix and that $\vect{\pi}=(\pi_A , \pi_G, \pi_C , \pi_T)$ is the associated stationary distribution. After applying the quadratic transformation~\eqref{eq:Q_j},  it follows from~(A.1) in 
Section~A of our Online Supplementary Materials
that the transversion and transition rates in the rate matrix for site $j$ are
$\beta_j=c_j \beta (1 + d_j \beta)$,  
$\rho_{1,j}=c_j [ \rho_1 + d_j \{ \rho_1^2 - (\rho_1 - \beta)^2 \pi_R\}]$,
$\rho_{2,j}=c_j [ \rho_2 + d_j \{ \rho_2^2 - (\rho_2 - \beta)^2\pi_Y\}]$,
where $\pi_R = \pi_A + \pi_G$ and $\pi_Y = \pi_C + \pi_T$.


If we take the distribution at the root of the tree to be the vector $\vect{\pi}$ satisfying $\vect{\pi} \matr{Q} = \vect{0}\transpose$ then the resulting Markov process is stationary and the term $\Pr(\vect{Y}_j = \vect{y}_j | \matr{Q}_j, \tau, \vect{\ell})$ in the likelihood~\eqref{eq:likelihood_quash1} is given by
\begin{equation}\label{eq:likelihood_quash2}
\Pr(\vect{Y}_j = \vect{y}_j | \matr{Q}_j, \tau, \vect{\ell}) = \sum_{X} \pi_{X(0)} \prod_{\text{edges} \hspace*{3.0pt} b = (v,w)} p_{j, X(v), X(w)}(\ell_b).
\end{equation}
Here $v$ and $w$ are the vertices (nodes) at the two ends of edge $b$ with length $\ell_b$, $X(u)$ is the character at vertex $u$, $u=0$ denotes the root vertex and $P_j(\ell)=\{ p_{j,u,u'}(\ell) \}$ is the transition matrix associated with an edge of length $\ell$ at site $j$. The sum is over all functions $X$ from the vertices to $\Omega_K$ such that $X(u)$ matches the data $y_j(u)$ for all leaf vertices $u$. It can be computed efficiently using a \label{pg:rev3_6}post-order traversal of the tree called Felsenstein's pruning algorithm \citep[][]{Fel73} or the sum-product algorithm in the context of Bayesian networks. 

\subsection{\label{subsec:quash_biol}Biological interpretation}\label{pg:rev3_1}
A biological interpretation for LASH and QuASH models can be obtained by considering how the substitution process in each case might result from the combination of a process of point mutation and a process of selection, where point mutations become fixed in a population.

The fixation rates of point mutations vary across sites according to differences in their structural or functional importance. As a consequence, sites under high selective constraints typically admit fewer substitutions of any type. At a more granular level, different types of mutations at any particular site, that is, mutations between different pairs of nucleotides, may have different \emph{selective coefficients}. These measure the relative fitness of a particular allele (point mutation), with larger numbers indicating stronger selection for (or against) the allele and hence higher selective pressure. 

Interpretations of the Markov process arising from the LASH and QuASH transformations are best explained through their representation as jump processes. To this end, consider a baseline, stationary substitution process with rate matrix $\matr{Q}$ that represents the combined processes of point mutation and selection at a ``typical'' site. We can characterise the behaviour along an edge of the tree with rate matrix $\matr{Q}$ as a jump process, which spends an exponentially $\expo(-q_{u,u})$ distributed time in nucleotide $u$ before transitioning to another nucleotide $v \ne u$ with probability $- q_{u,v} / q_{u,u}$. The LASH model applies the scaling $\matr{Q}_j = c_j \matr{Q}$ at site $j$. The resulting jump process retains the same jump probabilities as the ``typical'' site but the exponential dwell time in nucleotide $u$ now has rate parameter $-c_j q_{u,u}$. Biologically, we would expect qualitatively equivalent behaviour if each site evolved according to a common process of point mutation, with a site-specific fixation rate that was shared by all mutations at that site. In other words, we could regard the LASH model as allowing for differences in overall selective constraints across sites, but not for any heterogeneity in the site-specific selective coefficients for different types of point mutation.

Under the QuASH transformation, $\matr{Q}_j = c_j \matr{Q} - c_j d_j \matr{Q}^2$, and so, like the LASH model, the QuASH model has a site-specific parameter $c_j$ which allows for variation in the overall rate of evolution across sites. However, as a result of the transformation, the process at site $j$ now spends an exponentially $\expo(-c_j q_{u,u} + c_j d_j q_{u,u}^2 + c_j d_j K_u)$ distributed time in nucleotide $u$, where $K_u$ is the dot product of row $u$ and column $u$ of $\matr{Q}$ with $q_{u,u}$ removed. \label{pg:rev2v2_1}Compared to the mean in the baseline process $\matr{Q}$, the mean dwell time could have gone up for some nucleotides, and down for others because, given coefficients $c_j$ and $d_j$, it is possible that 
\begin{equation*}
-c_j q_{u,u} + c_j d_j q_{u,u}^2 + c_j d_j K_u > - q_{u,u}
\end{equation*}
for some $u \in \Omega_4$ whilst
\begin{equation*}
-c_j q_{v,v} + c_j d_j q_{v,v}^2 + c_j d_j K_v < - q_{v,v} 
\end{equation*}
for other $v \in \Omega_4$ where $v \ne u$; see our Online Supplementary Materials for a numerical example. Similarly, the probabilities of subsequent transition into nucleotide $v \ne u$ are no longer equal to $- q_{u,v} / q_{u,u}$. Of course, this is inevitable because the stationary distribution of $\matr{Q}_j$ is the same as that of $\matr{Q}$. Therefore, when compared to the process at a typical site, if dwell times at site $j$ are longer for nucleotide~$u$ and shorter for nucleotide~$v$, this must be compensated by smaller jump probabilities into nucleotide~$u$ and higher jump probabilities into nucleotide~$v$. Biologically, we can interpret the joint effect of the QuASH transformation on the dwell times and jump probabilities as representing the effects of heterogeneity in the selective coefficients for different types of point mutation. \label{pg:rev2v2_2b}Specifically, substitutions with high (advantageous) selective coefficients are rare, but if they do occur, then they persist in the population for a long time. This would be represented in the substitution process by smaller jump probabilities into the advantageous nucleotide but longer dwell times. In contrast, mutations with selective coefficients close to neutrality arise and are fixed more frequently, but can quickly be replaced. This would be represented in the substitution process by larger jump probabilities into the nucleotide in question and then short dwell times. Therefore, whilst we can interpret both the LASH and QuASH models as allowing for across-site variation in the overall selective constraints, only the QuASH model allows for across-site heterogeneity in selective behaviours.

\subsection{\label{subsec:random_effects}Random Effect Distribution}
We model the coefficients $\vect{d}=(d_1,\ldots,d_M)\transpose$ involved in the second order term of the quadratic transformation~\eqref{eq:Q_j} as conditionally i.i.d. with $d_j | \matr{Q}, \beta \sim \mathcal{F}(\beta)$ for some unknown hyperparameter $\beta$. As explained earlier in this section, the distribution $\mathcal{F}$ has support on $(l(\matr{Q}),u(\matr{Q}))$ where $l(\matr{Q})$ is nonpositive but assumed finite whilst $u(\matr{Q})$ is nonnegative but can be infinite. This means the interval $(l(\matr{Q}),u(\matr{Q}))$ can be finite or semi-infinite. In order to handle the two cases in a consistent fashion, we construct the distribution of $d_j$ through a shifted, piecewise power transformation of a Beta random variable
\begin{equation*}
d_j = 
\begin{cases}
l(\matr{Q})+w(\matr{Q})\left(1 - b_j^{1/w(\matr{Q})}\right), \quad &\text{if $u(\matr{Q})$ is finite,}\\
l(\matr{Q}) - \log b_j,                              &\text{otherwise,}
\end{cases}
\end{equation*}
where $w(\matr{Q})=u(\matr{Q})-l(\matr{Q})$; $b_j | \matr{Q}, \beta \sim \bet [
\beta+a(\matr{Q}), \beta\{b(\matr{Q})-1\}+1 ]$; and $\beta > 0$ is unknown. The terms $a(\matr{Q})$
and $b(\matr{Q})$ depend on the baseline rate matrix $\matr{Q}$ through
$a(\matr{Q})=1/w(\matr{Q})$ if $u(\matr{Q})$ is finite and $a(\matr{Q})=0$ otherwise, and
$b(\matr{Q})=\{w(\matr{Q})/u(\matr{Q})\}^{w(\matr{Q})}$ if $u(\matr{Q})$ is finite and $b(\matr{Q})=e^{-l(\matr{Q})}$ otherwise.
This choice ensures that the mode of the distribution is zero, with finite probability density, and that the density of $d_j$ decays smoothly to zero at its end points, except in the case where $l(\matr{Q})=0$ or $u(\matr{Q})=0$. In the special case when $l(\matr{Q})=0$ and $u(\matr{Q})$ is infinite, the conditional distribution of $d_j$ reduces to the $\expo(\beta)$ distribution. By \label{pg:rev3_7}centering the distribution on zero, we encourage shrinkage towards the nested LASH model with all $d_j = 0$. Although it may appear more natural to set the mean or median, rather than the mode, to zero, since the lower or upper end points of the support can be equal to zero, this is not possible in the general case. 

\begin{figure}[ht]
\centering
\includegraphics[scale=0.85]{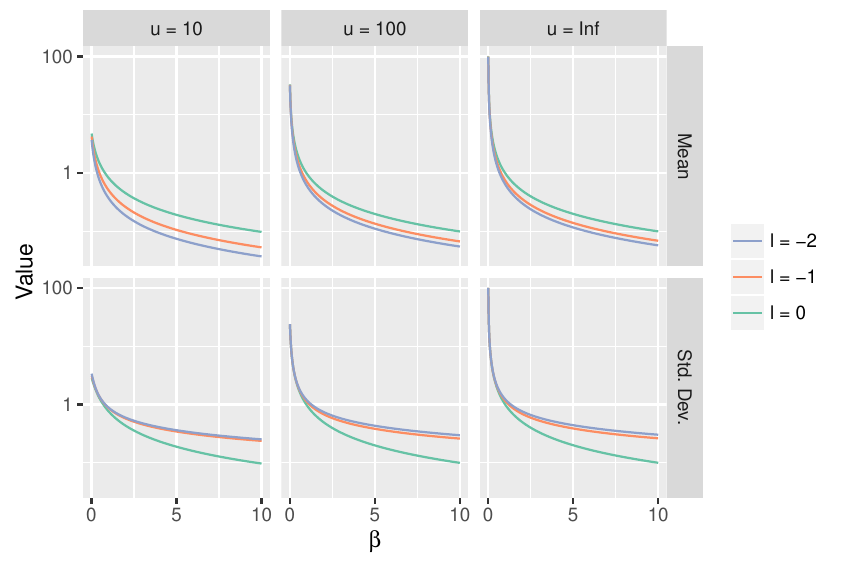}
\caption{Conditional mean and standard deviation of $d_j$ given $\beta$, and bounds $l$ and $u$, plotted with a log-scale on the $y$-axis.}
\label{fig:dj_prior}
\end{figure}

The hyperparameter $\beta$ can be assigned any prior with support on the positive real line. The dependence of the marginal prior for $d_j$ on that for $\beta$ and the parameters of the baseline rate matrix $\matr{Q}$ is complex. However, closed form expressions for the conditional expectation and variance of $d_j$ given $\beta$, and bounds $l$ and $u$ can be computed and are given in 
our Online Supplementary Materials. 
For various values of $l$ and $u$ spanning the range inferred in analyses of real data, Figure~\ref{fig:dj_prior} plots the conditional mean and standard deviation as a function of $\beta$. Clearly as $\beta$ gets large, the distribution of $d_j$ tends towards a point mass at zero and we recover a simple LASH model. However, as $\beta$ approaches zero, the mean and standard deviation both become large. Therefore we can allow more heterogeneity across sites by giving $\beta$ a prior which assigns reasonable density around zero.

\section{\label{sec:ns-quash}Non-stationary models}
The transformations characterising LASH and QuASH models allow across-site variation in the overall magnitude of the instantaneous rates of change and, for QuASH models, their relative sizes. However, the models discussed so far have been homogeneous across branches, with a single baseline rate matrix $\matr{Q}$ applying to the whole tree. Furthermore, the linear and quadratic transformations~\eqref{eq:Q_j_lin} and~\eqref{eq:Q_j} preserve the stationary distribution $\vect{\pi}$ of $\matr{Q}$. Therefore if the distribution at the root of the tree $\vect{\pi}_{(0)}$ is equal to $\vect{\pi}$, then the resulting Markov process will assume the same stationary distribution at all sites. These models cannot, therefore, explain the heterogeneities in sequence composition that are commonly observed in experimental data, either across taxa or across sites. As explained in Section~\ref{sec:intro}, the resulting model misspecification can lead to misleading phylogenetic inferences.

Non-stationary models for sequence evolution can account for differences in composition across taxa by allowing the probability of being in each state (e.g. each nucleotide for DNA data) to change over time. Typically this is achieved by permitting step changes in the theoretical stationary distribution at different points on the tree. Although these changes do not have to occur at speciation events \citep[e.g.][]{BL06}, this assumption is often made \citep[e.g.][]{YR95,Fos04,HNBWE14,Che16} and we retain it here for simplicity of notation. In general, therefore, consider a rooted topology $\tau$ with $B$ branches and a model that assumes a distribution $\vect{\pi}_{(0)}$ at the root of the tree, with the processes on the other branches governed by normalised rate matrices $\matr{Q}_{(1)},\ldots,\matr{Q}_{(B)}$, with associated theoretical stationary distributions $\vect{\pi}_{(1)},\ldots,\vect{\pi}_{(B)}$. To achieve non-stationarity we need $\vect{\pi}_{(b)} \ne \vect{\pi}_{(0)}$ for at least one $b \in \{1,\ldots,B\}$, however for some\label{pg:rev3_8} distinct branches $(b, b')$, $b \ne b'$, we might fix $\vect{\pi}_{(b)}$ to be equal to $\vect{\pi}_{(b')}$.

Extending the LASH and QuASH transformations to non-stationary models of this form, the rate matrix for site $j$ on branch $b$ is given by
\begin{equation*}
\matr{Q}_{b,j} = c_j \matr{Q}_{(b)} - c_j d_j \matr{Q}_{(b)}^2
\end{equation*}
where $c_j \in (0, \infty)$, whilst $d_j=0$ for LASH models and $d_j
\in (l, u)$ for QuASH models. In the latter case, the limits 
depend on all the $\matr{Q}_{(b)}$, with
$l = \max \{ l(\matr{Q}_{(b)}): b = 1,\ldots, B\}$ and $u = \min \{ u(\matr{Q}_{(b)}): b = 1,\ldots, B\}$,
where $l(\cdot)$ and $u(\cdot)$ are as in~\eqref{eq:lower} and~\eqref{eq:upper} respectively. This ensures that all the resulting $\matr{Q}_{b,j}$ are valid rate matrices. The likelihood expressions~\eqref{eq:likelihood_quash1} and~\eqref{eq:likelihood_quash2} for stationary QuASH models can now be modified to give
\begin{equation}\label{eq:likelihood_nsquash1}
p(\matr{y} | \matr{Q}_{(1)},\ldots,\matr{Q}_{(B)},\vect{\pi}_{(0)}, \tau, \vect{\ell}, \alpha, \beta) \simeq \prod_{j=1}^M \frac{1}{K_c K_d} \sum_{a=1}^{K_c} \sum_{a'=1}^{K_d} \Pr(\vect{Y}_j = \vect{y}_j | \matr{Q}_{1,j},\ldots,\matr{Q}_{B,j},\vect{\pi}_{(0)}, \tau, \vect{\ell})
\end{equation}
where $\matr{Q}_{b,j}$ is a \label{pg:rev3_9}function of $\left\{ z_{c,a}(\alpha), z_{d,a'}(\matr{Q}_{(1)},\ldots,\matr{Q}_{(B)},\beta), \matr{Q}_{(b)} \right\}$ and
\begin{equation}\label{eq:likelihood_nsquash2}
\Pr(\vect{Y}_j = \vect{y}_j | \matr{Q}_{1,j},\ldots,\matr{Q}_{B,j},\vect{\pi}_{(0)}, \tau, \vect{\ell}) = \sum_{X} \pi_{(0),X(0)} \prod_{\text{edges} \hspace*{3.0pt} b = (v,w)} p_{b, j, X(v), X(w)}(\ell_b)
\end{equation}
in which $P_{b,j}(\ell_b) = \{ p_{b, j, h,i}(\ell_b) \} = \exp (\ell_b \matr{Q}_{b,j})$ is the transition matrix associated with edge $b$, of length $\ell_b$, and site $j$.

\label{pg:rev1_1a}By definition, non-stationary models can allow heterogeneities in sequence composition across taxa. \label{pg:rev3_11}Consider, for example, a simple non-stationary model which allows a single step change in the stationary distribution at the root of the tree \label{pg:rev3_10}\citep[e.g.][Chapter 4]{KVR15,Che16}. In a site-homogeneous version of this model, a single rate matrix $\matr{Q}_{(1)}$, with associated stationary distribution $\vect{\pi}_{(1)} \not \equiv \vect{\pi}_{(0)}$, applies to all branches of the tree. In this case, the marginal distributions at the leaves depend on how long the process has had to move away from the distribution $\vect{\pi}_{(0)}$ at the root and converge towards $\vect{\pi}_{(1)}$ before reaching the tips of the pendant edges. In non-clock trees, where the leaf depths vary across taxa, this allows variation in the corresponding marginal distribution. Similarly in more complex models where there is more than one step change in the stationary distribution, the marginal distribution will vary due to differences in both the leaf depths (for non-clock trees) and differences in the sets of $\matr{Q}_{(b)}$ matrices on the evolutionary paths for different taxa.

Although more subtle, LASH and QuASH extensions of these non-stationary models additionally allow heterogeneity between sites in the across-taxa variation. For instance, consider the LASH or QuASH extension of the simple non-stationary model described above, regarding $\matr{Q}_{(1)}$ as the baseline rate matrix and denoting by $\matr{Q}_{1,j}$ the rate matrix associated with site $j$. If $\lambda$ is an eigenvalue of $\matr{Q}_{(1)}$, it follows immediately from~\eqref{eq:Q_j} that $c_j \lambda - c_j d_j \lambda^2$ is an eigenvalue of $\matr{Q}_{1,j}$, with $d_j=0$ for LASH models. Denote by $\lambda_{j,1},\lambda_{j,2},\ldots,\lambda_{j,K}$ the eigenvalues of $\matr{Q}_{1,j}$ ordered such that $\lambda_{j,1} = 0 > \mathrm{Re}(\lambda_{j,2}) \ge \mathrm{Re}(\lambda_{j,3}) \ge \cdots \ge \mathrm{Re}(\lambda_{j,K})$, where $\mathrm{Re}(\lambda)$ denotes the real part of the complex number $\lambda$. Under this model, it can be shown that $\matr{P}_j(\ell) = \vect{1} \vect{\pi}_{(1)} + O(e^{-\nu_j \ell})$ as $\ell \to \infty$ where $\vect{1}$ is a length $K$ column vector of 1s and $\nu_j = - \mathrm{Re}(\lambda_{j,2})$; see, for example, \citet{Kij97}, Chapter 4. It follows that at sites for which $\nu_j$ is large, the rate of convergence towards the stationary distribution $\vect{\pi}_{(1)}$ associated with $\matr{Q}_{(1)}$ will be fast, giving rise to marginal distributions at the leaves of the tree that resemble $\vect{\pi}_{(1)}$, especially for those taxa whose leaf depth is large. In contrast, at sites for which $\nu_j$ is small, the rate of convergence will be slow, leading to marginal distributions at the leaves that are closer to the distribution at the root $\vect{\pi}_{(0)}$. Again, this will be more pronounced for taxa with a small associated leaf depth. Although LASH and QuASH models both allow this kind of behaviour, in QuASH models it is managed more flexibly by two parameters, rather than one. Further, as discussed in Sections~\ref{subsec:quash_props} and \ref{subsec:quash_biol}, only the QuASH mapping allows the ratios of the instantaneous rates of change, and hence transition patterns, to vary across sites.

In the application in Section~\ref{subsec:apps_nonstat}, we focus on the HB model \citep[][]{HNBWE14} where each branch of the tree has its own reversible rate matrix $\matr{Q}_{(b)}$ which factorises into a composition vector $\vect{\pi}_{(b)}$ and a set of exchangeability parameters $\vect{\rho}$ that are assumed constant across the tree. We use the formulation of the model from \citet{WHCNBE15} in which the composition vector on the root edge of the underlying unrooted topology is the same as that at the root of the tree $\vect{\pi}_{(0)}$. To allow information to be shared between branches, the composition vectors $\{ \vect{\pi}_{(b)} \}$ are positively correlated \textit{a priori}. Full details can be found in the description of Prior B in \citet{HNBWE14} but, briefly, a greater exchange of information between neighbouring branches is admitted by adopting a first order autoregressive structure in which the composition vector on branch $b$ is conditionally independent of the composition vectors on all non-descendant branches given its parent.

\section{\label{sec:inference}Posterior inference via MCMC}
Let $\vect{\theta}$ represent the parameters of the distribution at the root of the tree and the set of baseline rate matrices. \label{pg:rev3_12}For a given tree $\tau$ and set of branch lengths $\vect{\ell}$, these parameters would be common to a site-homogeneous model and its LASH and QuASH extensions. For example, $\vect{\theta}=\{ \vect{\pi}, \vect{\rho} \}$ for a simple, stationary QuASH model based on a reversible rate matrix, or $\vect{\theta}=\{ \vect{\pi}_{(0)},\ldots,\vect{\pi}_{(B-2)}, \vect{\rho} \}$ for the HB variant. The joint posterior distribution for all unknowns is then
\begin{equation*}
p(\vect{\theta}, \tau, \vect{\ell}, \alpha, \beta | \matr{y}) \propto p(\matr{y} | \vect{\theta}, \tau, \vect{\ell}, \alpha, \beta)\, p(\vect{\theta}, \tau, \vect{\ell}, \alpha, \beta)\label{pg:rev3_13}
\end{equation*}
where the likelihood function $p(\matr{y} | \vect{\theta}, \tau, \vect{\ell}, \alpha, \beta)$ was given in~\eqref{eq:likelihood_quash1} and~\eqref{eq:likelihood_quash2} for a simple, stationary QuASH model, or in~\eqref{eq:likelihood_nsquash1} and~\eqref{eq:likelihood_nsquash2} for a non-stationary QuASH model.

Irrespective of the choice of prior distribution $p(\vect{\theta}, \tau, \vect{\ell}, \alpha, \beta)$, the posterior is analytically intractable. We therefore build up a numerical approximation using a Metropolis within Gibbs sampling scheme which iterates through a series of updates for each unknown. Real valued parameters, such as branch lengths $\vect{\ell}$, can be updated using standard proposal distributions, for example Gaussian random walks on the log-scale. In QuASH models whose likelihood is invariant to the root position, $\tau$ represents an unrooted topology which can be updated using standard topological moves such as nearest neighbour interchange (NNI) and subtree prune and regraft (SPR); see, for example, \citet{mrbayes}. For QuASH models whose likelihood depends on the root position, $\tau$ represents a rooted topology and so proposals which attempt to move the root are also required. In the applications in Section~\ref{sec:apps}, for example, we consider the QuASH variant of the HB model and employ the NNI, SPR and root moves described in \citet{HNBWE14}. These topological moves are complicated by the step changes in the theoretical stationary distribution which characterise the HB model. As there is \label{pg:rev3_14}a different composition vector associated with each branch of the underlying unrooted topology, topological moves include modifications to the composition vectors, as well as branch lengths, for the edges \label{pg:rev3_15}whose local interpretation changes under the proposed new topology. To generate such proposals, we can, for example, propose the new composition vectors using a distribution centred at the composition on a neighbouring branch; see \citeauthor{HNBWE14} for full details. The MCMC inferential procedures are programmed in Java. A software implementation can be found in the Online Supplementary Material.

\section{\label{sec:apps}Applications}
\label{pg:rev3_16}A controversial issue in evolutionary biology is the deep structure of the \emph{tree of life}, including the relationships among Bacteria, Archaea and eukaryotes, the three main cellular domains. \label{pg:rev1_4}The balance of evidence favours endosymbiotic hypotheses for the origin of eukaryotes, involving symbiosis between a bacterial endosymbiont (the mitochondrion) and some kind of host cell \citep[][]{MGZ15}. \citet{WKW90} proposed that this host cell was part of an independently-branching third domain of life, distinct from Archaea and Bacteria. This is often referred to as the \emph{three domains hypothesis}. On the basis of analyses involving previously unsequenced taxa and more sophisticated evolutionary models \citep[][]{WFCE13}, an alternative view -- the \emph{eocyte hypothesis} -- has gained considerable support over recent years. According to this conjecture, the host for the mitochondrial endosymbiont was a fully-fledged Archaeon. In addition to uncertainty surrounding the unrooted topology of the tree of life, opinion is also divided on the position of its root. Under the two leading hypotheses, it is either placed on the bacterial branch \citep[][]{GKDTBBMPDO89,IKHOM89} or, with fewer proponents, within the Bacteria \citep[][]{Cav06,LSHS09}.  

In this section we consider applications to biological data sets that address these controversial questions. In Section~\ref{subsec:apps_stat} we analyse a concatenated alignment of small and large subunit ribosomal RNAs (SSU and LSU rRNAs) sampled from across the tree of life. After alignment using MUSCLE \citep[][]{Edg04} and editing to remove poorly-aligning regions, $M=1734$ sites on $N=36$ species remained. We consider three models: (S1) a stationary, reversible TN93 model, (S2) the LASH-variant of S1 and (S3) the QuASH-variant of S1. Models that are stationary and reversible give rise to likelihood functions that are invariant to the position of the root, and so these analyses only allow inference of the unrooted topology. In Section~\ref{subsec:apps_nonstat} we therefore consider three non-stationary models which also allow us to learn about the root position: (NS1) the HB model with TN93 exchangeability parameters; (NS2) the LASH-variant of NS1 and (NS3) the QuASH-variant of NS1. Inference via MCMC is substantially slower for the HB model and so, for computational tractability, we consider a smaller data set with $M=1481$ sites and only $N=16$ taxa. Further discussion on the scalability of our model-fitting procedures can be found in the Appendix.

In all analyses, mixing and convergence of the MCMC sampler was assessed by comparing the output from multiple chains, initialised at different starting points. In phylogenetics, mixing in tree-space can be problematic due to the low acceptance rates of topological moves. Therefore, in addition to considering the usual numerical and graphical diagnostic checks for continuous parameters, we also examined graphs based on relative cumulative split (Section~\ref{subsec:apps_stat}) or clade (Section~\ref{subsec:apps_nonstat}) frequencies of the chains over the course of the MCMC runs; see \citet{HNBWE14} for full details of these diagnostics. Here a \emph{split} refers to a bipartition of the taxa at the leaves of the tree into two disjoint sets, induced by cutting a branch. On a rooted tree, one of the partition subsets of any split is a \emph{clade} if all the taxa lie on the same side of the root. In biological terms, this corresponds to an ancestor and all its descendants.

\subsection{\label{subsec:apps_stat}Stationary TN93 model}
Based on our subjective assessments of the evolutionary process, for the parameters of the S1 model we chose independent gamma $\mathrm{Gam}(1, 1)$ priors for the two transition rates $\rho_1$ and $\rho_2$ and a flat Dirichlet $\mathscr{D}(1,1,1,1)$ prior for the stationary distribution $\vect{\pi}$ in the unnormalised rate matrix. \label{pg:rev3_17b}In keeping with experiences from the literature, our posterior inferences were robust against reasonable modifications to this prior specification \citep[][]{ZH04}. We also specified independent exponential $\expo(10)$ priors for the branch lengths $\vect{\ell}$ and a uniform prior over unrooted topologies $\tau$. This expresses the prior belief that a branch represents 0.1 substitutions per site, on average, along with prior indifference with regards to the unrooted topology. In models S2 and S3 we additionally assigned a gamma $\mathrm{Gam}(10, 10)$ prior to the shape parameter $\alpha$ in the random effects distribution for the rates $c_j$ and, in model S3, a gamma $\mathrm{Gam}(1, 1)$ prior to the parameter $\beta$ in the random effects distribution for the quadratic coefficients $d_j$ of the QuASH model. The latter distribution, with mean $\E(\beta)=1$ and coefficient of variation $\mathrm{CV}(\beta)=1$, was chosen to give reasonable support to values of $\beta$ near zero. As explained in Section~\ref{subsec:random_effects}, this choice makes the prior for the $d_j$ reasonably diffuse. In order to check sensitivity to the prior specification for $\beta$, we repeated the analysis with model S3 using priors that had the same mean but different coefficients of variation and different behaviour near zero: $\mathrm{Gam}(10, 10)$ ($CV(\beta) = 0.316$) and $\mathrm{Gam}(0.1, 0.1)$ ($CV(\beta) = 3.16$). The phylogenetic and posterior predictive inferences reported in this section were robust against these changes.

\label{pg:rev3_18}We refer to the output of each complete sweep through the Gibbs steps of our Metropolis within Gibbs samplers as a single draw from the posterior. For each model the MCMC algorithm outlined in Section~\ref{sec:inference} was used to generate at least $110K$ draws from the posterior, after a burn-in of $100K$ samples, thinning the remaining output to retain every 100-th iterate. The diagnostics checks described earlier gave no evidence of any lack of convergence.

\begin{figure}[!p]
\centering
\subfloat[][]{\label{fig:s_consensus_1}\includegraphics[scale=0.6]{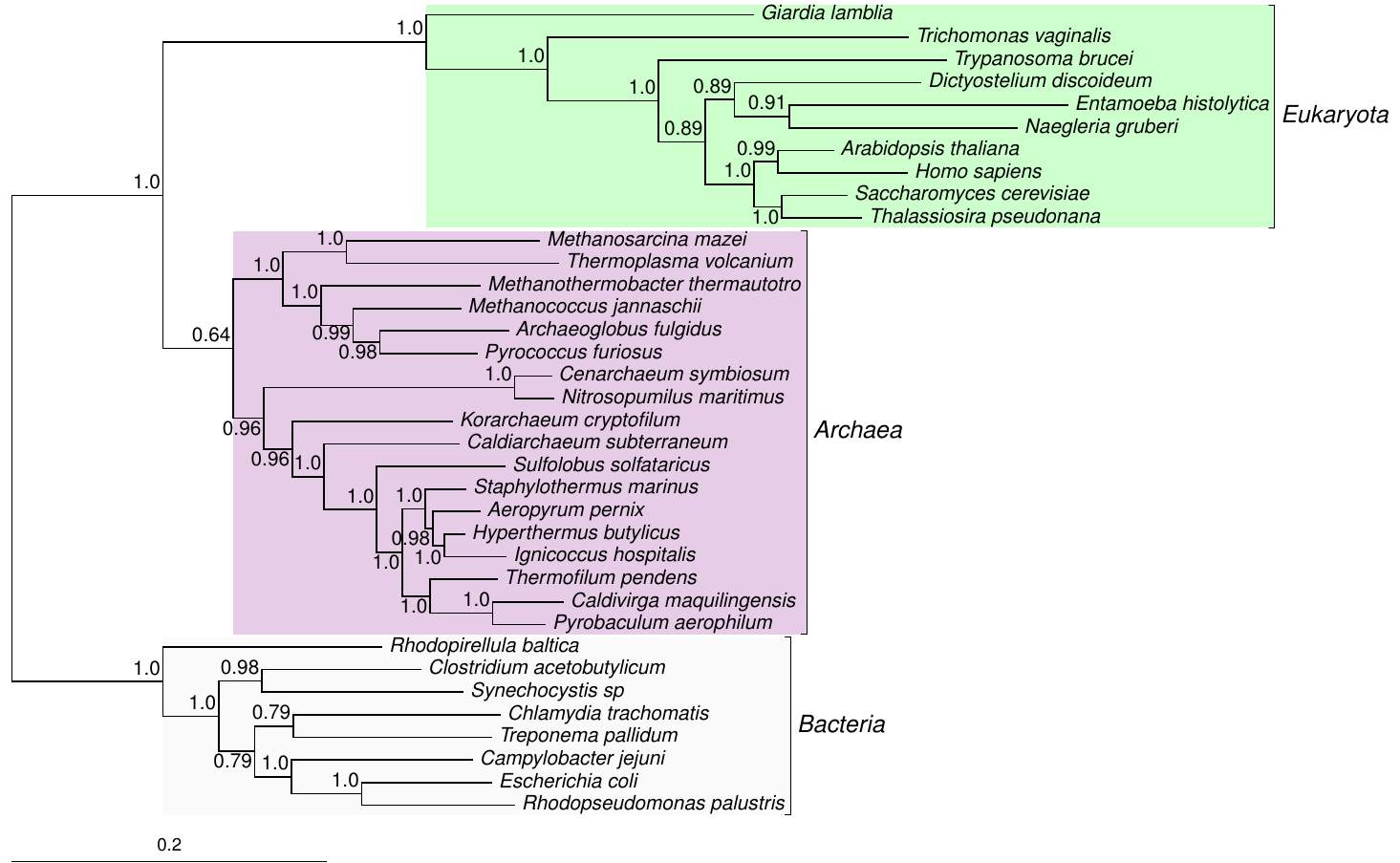}}\\[2cm]
\subfloat[][]{\label{fig:s_consensus_2}\includegraphics[scale=0.6]{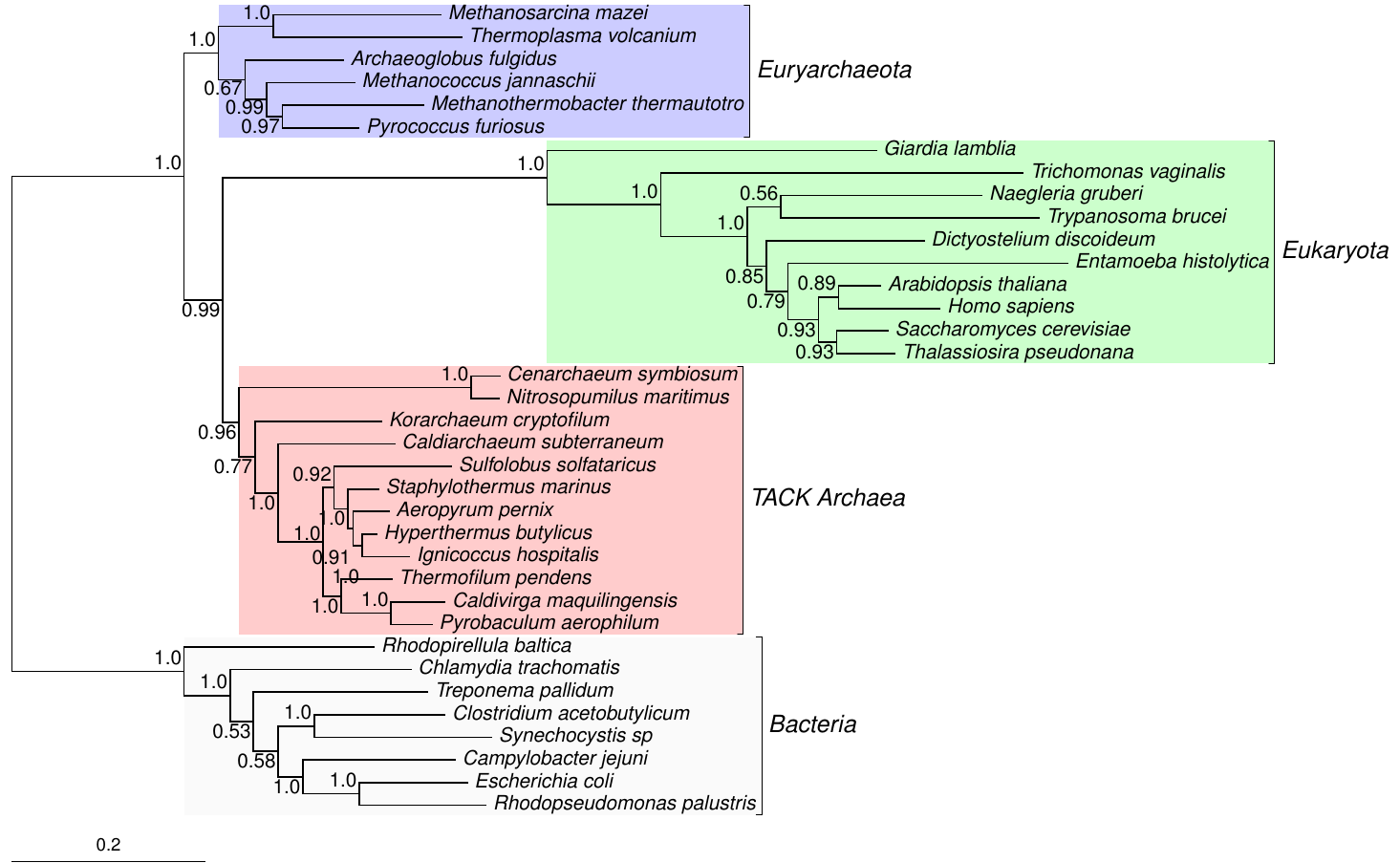}}
\end{figure}

\begin{figure}[!t]
\centering
\setcounter{figure}{2}
\ContinuedFloat
\subfloat[][]{\label{fig:s_consensus_3}\includegraphics[scale=0.6]{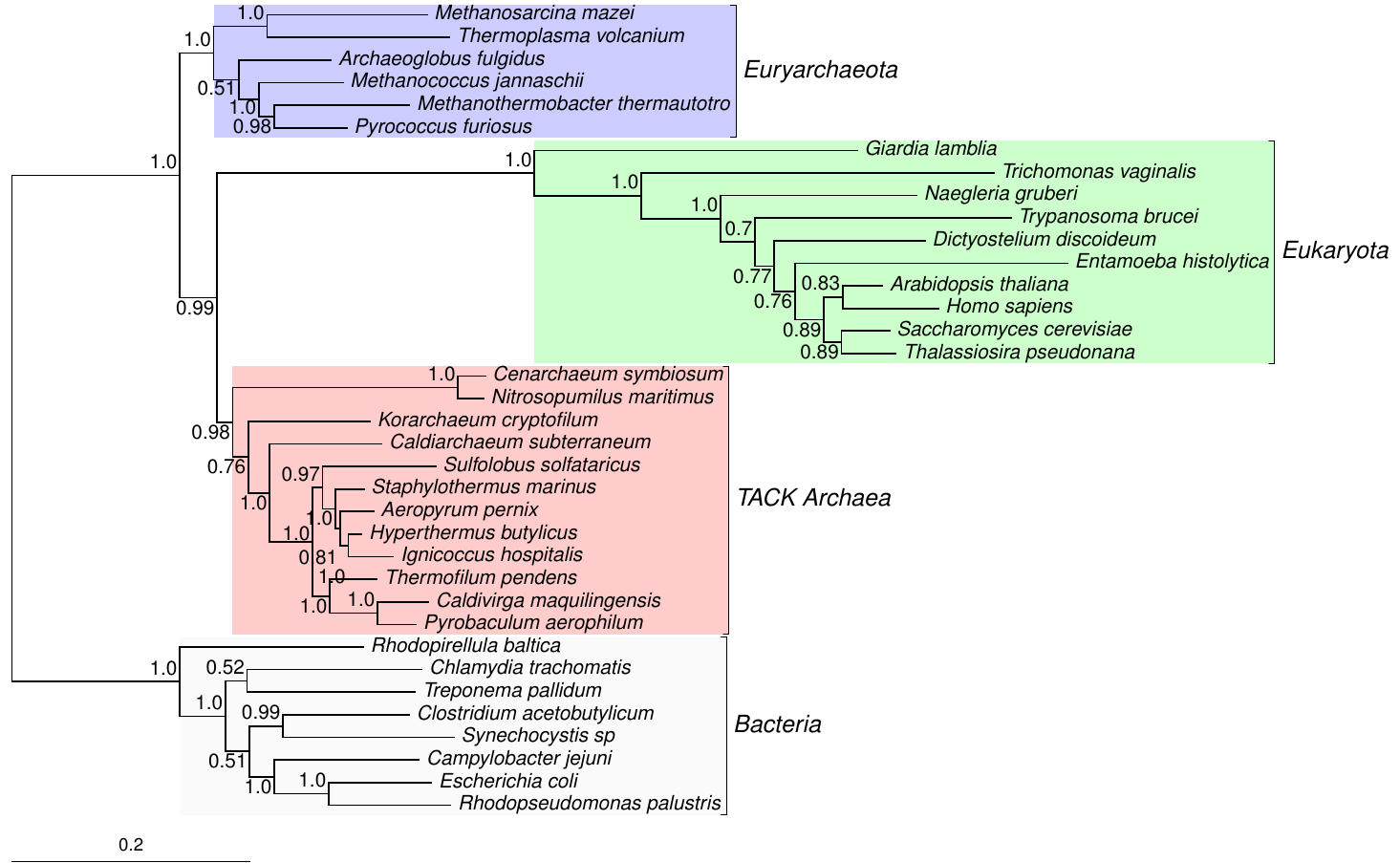}}
\caption{Majority rule consensus trees under models \protect\subref{fig:s_consensus_1} S1, \protect\subref{fig:s_consensus_2} S2 and \protect\subref{fig:s_consensus_3} S3. Numerical labels represent the posterior probability of the associated split. Branch lengths can be interpreted as the expected number of substitutions per site. Trees are unrooted but visualised with the root at the midpoint of the bacterial branch.}
\label{fig:s_consensus}
\end{figure}

\begin{figure}[!t]
\centering
\subfloat[][]{\label{fig:s_densbeta}\includegraphics[scale=0.85]{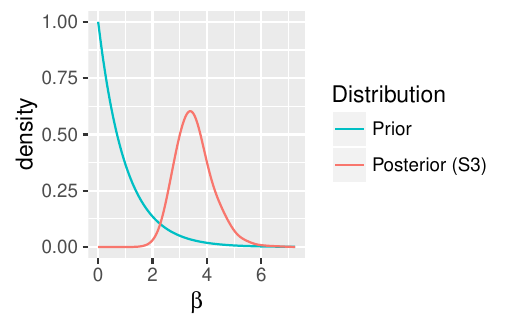}}
\subfloat[][]{\label{fig:ns_densbeta}\includegraphics[scale=0.85]{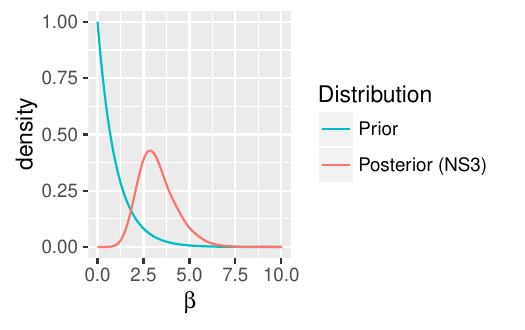}}\\
\subfloat[][]{\label{fig:s_densdj}\includegraphics[scale=0.85]{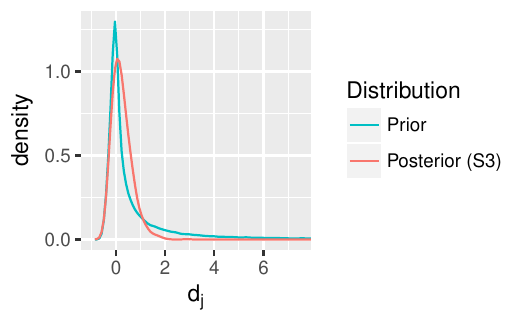}}
\subfloat[][]{\label{fig:ns_densdj}\includegraphics[scale=0.85]{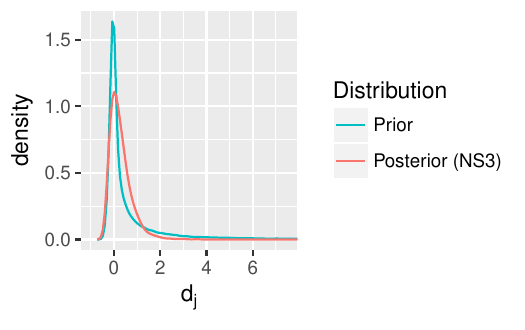}}\\
\caption{Top row: marginal prior and posterior densities for the unknown parameter $\beta$ in the random effect distribution for the quadratic coefficients $d_j$ under the \protect\subref{fig:s_densbeta} stationary model S3 and \protect\subref{fig:ns_densbeta} non-stationary model NS3. Bottom row: prior and posterior predictive distributions for $d_j$ at an additional site $j$ under the \protect\subref{fig:s_densdj} stationary model S3 and \protect\subref{fig:ns_densdj} non-stationary model NS3.}
\label{fig:densbeta}
\end{figure}

In phylogenetic inference, the majority-rule consensus tree is the most widely used summary of the posterior distribution over tree space. As a summary of a sample of trees, it includes only those splits which appear in over half of the samples \citep[][]{Bry03}, here representing those with posterior probability greater than 0.5. For the analyses under models S1--S3, the consensus trees are shown in Figure~\ref{fig:s_consensus} in which numerical labels represent the posterior probability of the associated split. To aid comparison, the trees are all visualised with the root on the bacterial branch. The consensus tree under S1 supports the three domains hypothesis, whilst models S2 and S3 yield eocyte trees, with eukaryotes emerging from within two archaeal clades: the Euryarchaeota and the TACK Archaea. As expected, there is a marked difference in our phylogenetic inferences as we move from the simple TN93 model (S1) to one which incorporates across-site rate heterogeneity. However, there is very little difference in the inferences obtained when extending the LASH model (S2) to the corresponding QuASH model (S3). Comparing the prior and posterior density for $\beta$ in Figure~\ref{fig:s_densbeta}, the posterior seems to support larger values for $\beta$ than the prior, which suggests a distribution for the quadratic coefficients $d_j$ that is more concentrated around zero. \label{pg:rev3_3a}Indeed this effect is borne out in Figure~\ref{fig:s_densdj} which shows that the prior predictive density for $d_j$ at an unobserved site $j$ has a much longer tail on the right than the corresponding posterior predictive density, all of whose mass is concentrated in a small neighbourhood around zero. The data do not, therefore, provide much evidence that the QuASH transformation is necessary given a model that already incorporates across-site rate heterogeneity.

\label{pg:rev2_3a}In order to compare the fit of models S1, S2 and S3, we use the framework of posterior predictive checks \citep[][]{GCSDVR13} in which the basic idea is to measure the extent to which a model captures some data summary of interest -- a so-called \emph{test statistic} -- by comparing its posterior predictive distribution to the value that was observed. Typically the posterior predictive distribution is approximated numerically based on an MCMC sample from the posterior of the unknowns in the model by simulating replicated data sets in one-to-one correspondence with the posterior draws. If the model is able to capture adequately the aspect of the data summarised through the test statistic, the observed value should look plausible under its posterior predictive distribution.

\begin{figure}[!t]
\centering
\subfloat[][]{\label{fig:s_postpred_bs}\includegraphics[scale=0.85]{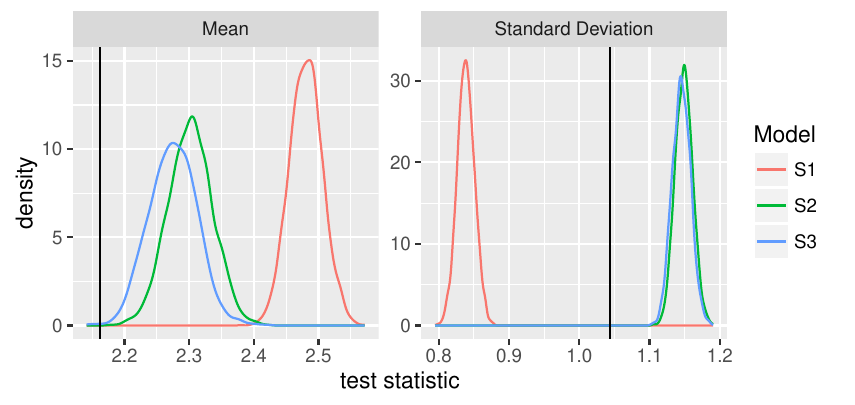}}\\
\subfloat[][]{\label{fig:ns_postpred_bs}\includegraphics[scale=0.85]{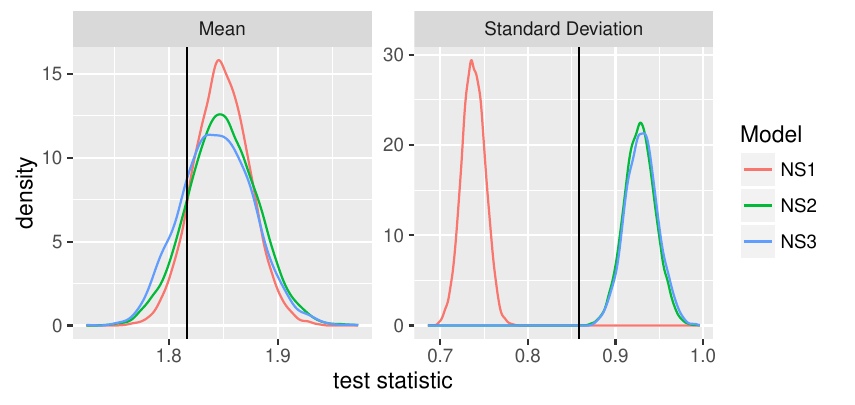}}
\caption{Posterior predictive densities for the mean and standard deviation of the number of distinct nucleotides per site in the analysis under the \protect\subref{fig:s_postpred_bs} stationary models S1--S3 and \protect\subref{fig:ns_postpred_bs} non-stationary models NS1--NS3. The observed values are indicated by vertical lines.}
\label{fig:postpred_bs}
\end{figure}

As explained in Section~\ref{sec:intro}, functional and structural constraints acting on a particular site can cause it to evolve very slowly. In such cases we are likely to see little or no variation in the character state at that column of the alignment. Therefore in fitting to the alignment-wide empirical compositions, models that do not allow variation in, at least, the rate of the evolutionary process across sites tend to overestimate the mean number of distinct nucleotides per column, and underestimate the associated standard deviation. Figure~\ref{fig:s_postpred_bs} shows the posterior predictive distribution for these test statistics obtained under models S1, S2 and S3, together with the observed values calculated from the alignment. \label{pg:rev2_3b}As expected, model S1 markedly overestimates the number of distinct nucleotides per site and underestimates the associated standard deviation. Whilst models S2 and S3 also overestimate the mean, the discrepancies are much less marked, with the QuASH-variant of the TN93 model (S3) being most compatible with the observed data. Interestingly, models S2 and S3 overestimate the standard deviation of the number of distinct nucleotides per site. It is possible that models allowing sequence composition to vary across sites would be required to adequately capture this feature.

\subsection{\label{subsec:apps_nonstat}Non-stationary HB model}
For the analyses using the non-stationary models NS1, NS2 and NS3, we adopted the prior distributions outlined in Section~\ref{subsec:apps_stat} for the two transition rates $\rho_1$ and $\rho_2$, the branch lengths $\vect{\ell}$ and the parameters $\alpha$ and $\beta$ in the random effects distributions for the linear and quadratic coefficients $c_j$ and $d_j$. As the HB model yields a likelihood function that depends on the position of the root, our topology $\tau$ is rooted. We assigned $\tau$ a prior according to the biologically-motivated Yule model of speciation, which generates a distribution in which near equal probability is assigned to root splits of all sizes: $1  \! :  \! (N-1)$, $2  \! :  \! (N-2)$, and so on \citep[][]{CHNBWE15}. For the composition vectors $\vect{\pi}_b$, $b=0,\ldots,B-2$, in the baseline rate matrix we used Prior B from \citet{HNBWE14}, choosing the hyperparameters representing the autoregressive coefficient and conditional variance to be $a=0.94$ and $b=0.31$ respectively. This specification was guided by simulations from the prior predictive distribution which suggested it led to a biologically plausible degree of heterogeneity in empirical sequence composition.

For each model the MCMC algorithm was used to generate at least $510K$ draws from the posterior, after a burn-in of $500K$ samples, thinning the remaining output to retain every 100-th iterate. The diagnostics checks described earlier gave no evidence of any lack of convergence.

\begin{figure}[!t]
\centering
\subfloat[][]{\label{fig:ns_consensus_1}\includegraphics[scale=0.6]{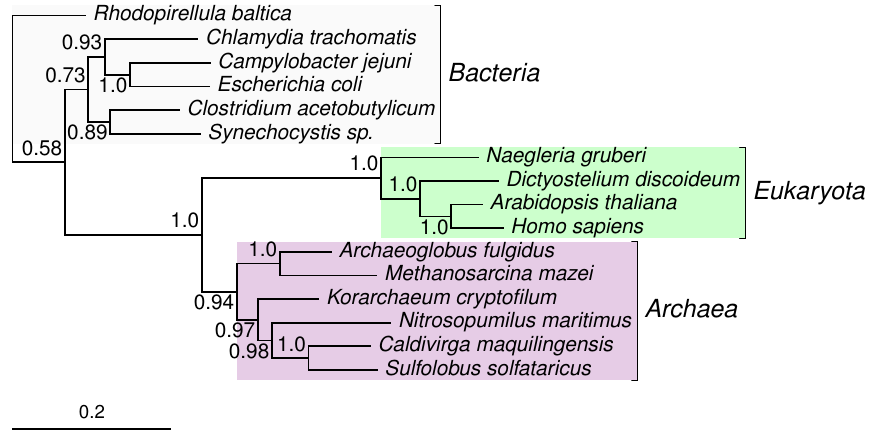}}\\
\subfloat[][]{\label{fig:ns_consensus_2}\includegraphics[scale=0.6]{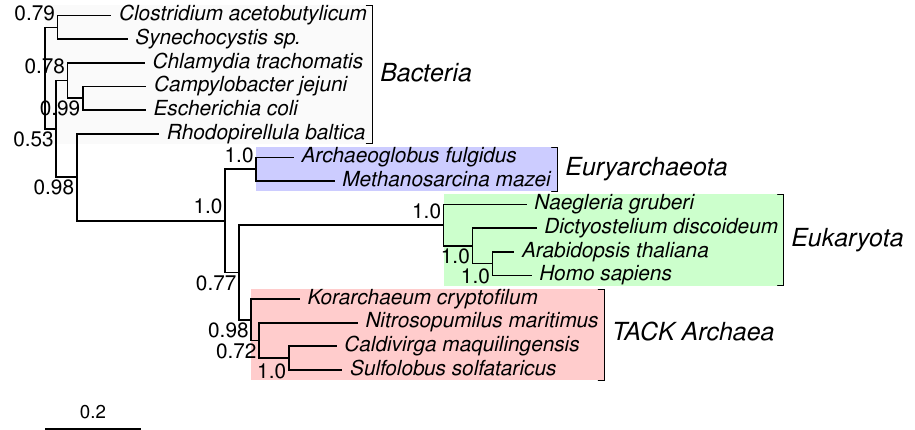}}\\
\subfloat[][]{\label{fig:ns_consensus_3}\includegraphics[scale=0.6]{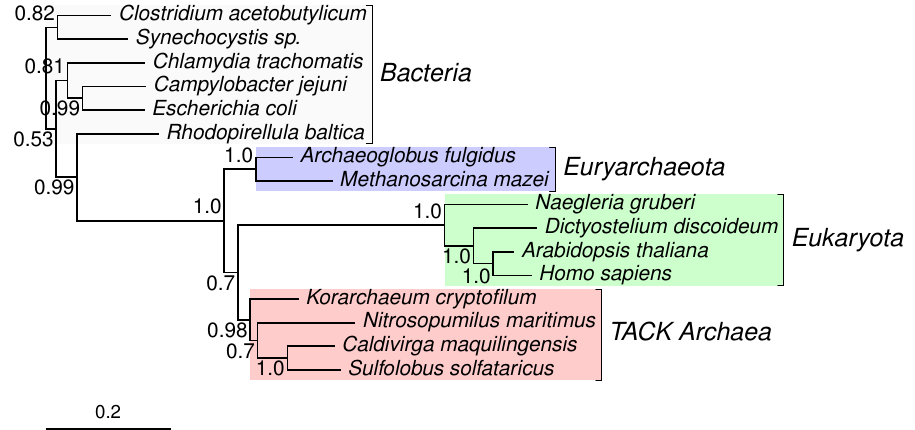}}
\caption{Rooted majority rule consensus trees under models \protect\subref{fig:ns_consensus_1} NS1, \protect\subref{fig:ns_consensus_2} NS2 and \protect\subref{fig:ns_consensus_3} NS3. Numerical labels represent the posterior probability of the associated clade. Branch lengths can be interpreted as the expected number of substitutions per site.}
\label{fig:ns_consensus}
\end{figure}

The rooted majority-rule consensus trees for each model are shown in Figure~\ref{fig:ns_consensus}. Our conclusions are consistent with those from Section~\ref{subsec:apps_stat}. Specifically, the model NS1 supports a three-domains tree whilst models NS2 and NS3 support very similar eocyte trees with, in this case, the same rooted topology. Although the site-homogeneous HB model (NS1) and the LASH and QuASH variants (NS2 and NS3) support different conclusions about the unrooted topology, they both suggest a root within the Bacteria. The marginal posterior distribution for root splits under the three models is summarised in Table~D.3 of our Online Supplementary Materials. Again, the differences between inferences under NS1 and NS2 are much more marked than those between NS2 and NS3. However, in all cases the posterior probability for a root within the Bacteria is 1.0.

The LASH and QuASH variants of the HB model allow sequence composition, as well as the overall rate of evolution, to vary across sites. Therefore we expect these models to be better equipped to capture the number of distinct nucleotides per site. Posterior predictive densities of the across-site mean and standard deviation are plotted in Figure~\ref{fig:ns_postpred_bs}. For the mean, all three models capture the observed statistic well, with the site-homogeneous model (NS1) offering slightly more support to larger values, as expected. \label{pg:rev2_3c}As in the analysis from Section~\ref{subsec:apps_stat}, the site-homogeneous model very markedly underestimates the standard deviation. The posterior predictive densities under the LASH (NS2) and QuASH (NS3) variants of the HB model are very similar. Although both overestimate the standard deviation, the observed statistic is more plausible than under the NS1 model, and the overestimation seems less marked than the corresponding comparison from Section~\ref{subsec:apps_stat}. The similarity in both phylogenetic and posterior predictive inferences under the LASH and QuASH models are consistent with the implications of the comparison between the prior and posterior in Figure~\ref{fig:densbeta}. \label{pg:rev3_3b}Figure~\ref{fig:ns_densbeta} shows the prior and posterior densities for $\beta$, whilst Figure~\ref{fig:ns_densdj} shows the prior and posterior predictive densities for the quadratic coefficient $d_j$ at an unobserved site $j$. As in the analysis of the stationary models, the posterior suggests a distribution for $d_j$ that is more concentrated around zero which suggests that the QuASH transformation adds only a small amount to a model in which linear across-site heterogeneity is already included.



\section{\label{sec:discussion}Discussion}
The introduction of across-site rate heterogeneity into substitution models for sequence evolution led to substantial improvements in model fit and the credibility of phylogenetic inferences. In practice, this feature was incorporated through a set of site-specific rates, modelled as random effects with unit mean gamma distribution, that linearly transformed a baseline rate matrix. Motivated by the advancement gained through this simple innovation, we considered a natural extension of the LASH model based on the incorporation of two sets of random effects, allowing a more flexible site-specific quadratic transformation of the baseline rate matrix. Biologically, this model makes fewer assumptions than the (nested) LASH model and allows for the effects of variation in the selective coefficients of different types of point mutation at a site, in addition to heterogeneity in overall selective constraints across sites. We derived properties of QuASH-transformed rate matrices, showing that they retain the stationary distribution of the underlying baseline matrix, and that the set of reversible rate matrices is closed under our quadratic transformation. In the context of a class of non-stationary models which permit step-changes in the theoretical stationary distribution at one or more points on the tree, we demonstrated that both the LASH and QuASH transformations lead to models which allow sequence composition to vary across sites as well as across taxa. This is due to different rates of convergence towards the theoretical stationary distributions at different sites. The QuASH-transformed, non-stationary models therefore provide a parsimonious means of allowing heterogeneity in sequence composition across both alignment dimensions.

We utilised our model and inferential procedures in two biological applications concerning the tree of life. In the first, we compared inferences under a stationary, reversible TN93 model, with those obtained under the LASH and QuASH extensions. In the second, to make computational inference manageable, we considered a smaller data set and compared inferences under a non-stationary HB model to those obtained under the LASH and QuASH variants. In both applications we found that the simpler site-homogeneous models supported the three domains hypothesis, with the Archaea, Bacteria and eukaryotes appearing as monophyletic groups. Conversely the more flexible LASH and QuASH models supported the eocyte hypothesis, with eukaryotes emerging from within a paraphyletic Archaea. \label{pg:rev3_20}The non-stationary models consistently placed the universal root within the Bacteria. The marked differences between inferences obtained under the site-homogeneous and LASH models are similar to other results reported in the literature \citep[][]{Yan96}. Both analyses suggested that only a small gain was achieved through the quadratic transformation once a linear mapping was in place. We have drawn similar conclusions from applications to several other data sets not reported here.

\label{pg:rev2_3d}Although our analyses have reinforced the importance of allowing heterogeneity in the rate of evolution across sites, it appears that only a modest benefit can be found by using a natural extension which exploits a quadratic transformation of the base rate matrix. This may be because the implications of heterogeneity across sites in the selective coefficients of different types of point mutation are difficult to detect from alignments of sequence data. \label{pg:rev2v2_2a}This might be particularly true of the ribosomal RNA sequences we analysed here, which are under strong selective constraints imposed both by the function of the molecule and by the physical interactions among sites that are separated in the primary sequence. However, in the context of non-stationary models, it is worth emphasising that even the LASH transformation generates models that allow heterogeneity in sequence composition across sites as well as across taxa. To our knowledge, this is a property that has gone unnoticed in the literature. Whilst a few, more mechanistic models have been proposed to offer this flexibility \citep[e.g.][]{BL08,JWRPJ14}, their complexity has made model-fitting computationally prohibitive. In contrast, non-stationary LASH and QuASH models provide a more parsimonious, data-driven alternative for which computational inference is substantially more straightforward. Our software implementation, described in the Appendix, provides a tool which allows practitioners to fit these models to their biological data sets.

\section*{References}
\bibliographystyle{plainnat}
\bibliography{journal_names_full,refs}

\section*{Appendix}\label{pg:rev2_1b}
A software implementation is available from

\centerline{\url{http://www.mas.ncl.ac.uk/~nseg4/QuASH/}}

The analyses in this paper were performed on a 2.40GHz Dell PowerEdge R410 server with two six-core Intel Xeon E5645 CPUs and 32GB RAM. When fitting the S2 and S3 models to the alignment from Section~\ref{subsec:apps_stat}, generating 500K MCMC samples took approximately 4 and 16 days, respectively. When fitting the more complex NS2 and NS3 models to the alignment from Section~\ref{subsec:apps_nonstat}, it took approximately 2.5 and 10 days, respectively, to generate 500K MCMC samples. In principle our software could be used to analyse alignments with any number of taxa and any number of sites. However, increasing the number of sites or the number of taxa increases run times. For example, for both sets of analyses detailed above, doubling the number of sites or the number of taxa roughly doubled the computational time. Clearly the size of the data sets that we can feasibly analyse is limited due to the computational complexity of the models considered. However, as demonstrated in Section~\ref{sec:apps} for the tree of life, by fitting more complex, biologically plausible models, even to relatively small data sets, we can challenge biological assumptions that would otherwise remain uncontested.

\section*{Acknowledgements}
SEH, TAW, SC and TME were supported by a European Research Council Advanced Investigator Award (ERC-2010-AdG-268701), and a Programme Grant from the Wellcome Trust (number 045404) to TME.

\end{document}